\documentclass[twocolumn,showpacs]{revtex4-1}
\usepackage{graphicx}
\usepackage{dcolumn}
\usepackage{epstopdf}
\graphicspath{{./pictures/}{./}}

\begin{document}
\title{Segmented scintillation detectors with silicon photomultiplier readout for measuring antiproton annihilations}
\author{A.~S\'ot\'er}
\email[]{anna.soter@mpq.mpg.de}
\affiliation{Max-Planck-Institut f\"ur Quantenoptik, Hans-Kopfermann-Strasse 1, D-85748 Garching, Germany}
\author{K.~Todoroki, T.~Kobayashi}
\affiliation{Department of Physics, University of Tokyo, 7-3-1 Hongo, Bunkyo-ku, Tokyo 113-0033, Japan}
\author{D.~Barna}
\affiliation{Department of Physics, University of Tokyo, 7-3-1 Hongo, Bunkyo-ku, Tokyo 113-0033, Japan}
\affiliation{Wigner Research Center of Physics, H-1525 Budapest, Hungary}
\author{D.~Horv\'ath}
\affiliation{Wigner Research Center of Physics, H-1525 Budapest, Hungary}
\author{M.~Hori}
\affiliation{Max-Planck-Institut f\"ur Quantenoptik, Hans-Kopfermann-Strasse 1, D-85748 Garching, Germany}
\affiliation{Department of Physics, University of Tokyo, 7-3-1 Hongo, Bunkyo-ku, Tokyo 113-0033, Japan}

\begin{abstract}
The Atomic Spectroscopy and Collisions Using Slow Antiprotons (ASACUSA) experiment at the Antiproton Decelerator (AD) facility of CERN constructed segmented scintillators to detect and track the charged pions which emerge from antiproton annihilations in a future superconducting radiofrequency Paul trap for antiprotons. 
A system of 541 cast and extruded scintillator bars were arranged in 11 detector modules which provided a spatial resolution of 17 mm. Green wavelength-shifting fibers were embedded in the scintillators, and read out by silicon photomultipliers which had a sensitive area of $1\times 1$ mm$^2$.
The photoelectron yields of various scintillator configurations were measured using a negative pion beam of  momentum $p\approx1$ GeV/c. Various fibers and silicon photomultipliers, fiber end terminations, and couplings between the fibers and scintillators were compared. The detectors were also tested using the antiproton beam of the AD. Nonlinear effects due to the saturation of the silicon photomultiplier were seen at high annihilation rates of the antiprotons.
\end{abstract}
\maketitle 

\section{Introduction}
 
The Atomic Spectroscopy and Collisions Using Slow Antiprotons (ASACUSA) experiment at the Antiproton Decelerator (AD) facility of CERN has recently developed position-sensitive detectors to measure and track charged pions emerging from antiproton annihilations. These segmented plastic scintillators will be used to detect annihilations that occur inside a radiofrequency Paul trap for antiprotons. They were also employed in a recent experiment which attempted to measure the annihilation cross-sections of low-energy (130 keV) antiprotons on various target foils \cite{pbaranni}. In this paper we describe the design and manufacture of these detectors, and beam tests carried out using negative pion ($\pi^-$) and antiproton beams provided by the CERN Proton Synchrotron (PS) and AD facilities.

Depending on the atomic nucleus on which an antiproton annihilates, on average 3--4 charged pions emerge with a mean kinetic energy of a few hundred MeV \cite{horicherenkov}. By tracking the trajectory of at least two charged pions, the initial vertex where the annihilation occurred can be reconstructed. This technique forms the basis of many experiments now carried out at the AD. Trackers based on silicon strip detectors \cite{athenaNature,alphaNature} or scintillation fibers \cite{atrap2002} have been previously used to detect the formation of antihydrogen, or to count the number of antiprotons stored in traps. Precision laser spectroscopy of antiprotonic helium atoms have been carried out by using acrylic Cherenkov counters to detect the timing of antiproton annihilations occurring in a cryogenic helium target \cite{mhori2001,nature_2011}.

Recently, extruded scintillators \cite{pla-dalmau3,pla-dalmau2001,pla-dalmau2003,pla-dalmau2} read out by wavelength-shifting (WLS) fibers \cite{dyshkant} and silicon photomultipliers (SiPMs) \cite{buzhan2003,MPPCdevelopment,renker2006} were used as tracker detectors in long-baseline neutrino experiments \cite{t2k_experiment,izmaylov}. They may also be used in electromagnetic calorimeters in future collider experiments \cite{LC,lc2}. Compared to conventional scintillators which are read out by photomultiplier tubes, the SiPM-based design has important advantages which include low cost, compact size, and insensitivity to magnetic fields. There are, however, currently numerous disadvantages such as the comparably small active area, gain, and dynamic range of the SiPMs; their high dark current and temperature sensitivity \cite{MPPCdevelopment,MPPCperformance}; and the narrow range of bias voltages in which the SiPMs can be operated in an optimal way.  In this paper, we systematically studied the response of the scintillator against individual hits of minimum-ionizing pions. It was particularly important to achieve a sufficient photoelectron yield, above the backgrounds of dark current and environmental noise.

This paper is organized in the following way. In Sect.~\ref{construction}, we describe the construction of the detector, with details of the SiPMs, scintillators, and WLS fibers. In Sect.~\ref{T9}, measurements carried out at using a 1-GeV/c pion beam provided by the CERN PS are described. We compared the performance of cast and extruded scintillators, fibers of various diameters, two types of SiPMs, and various methods to optically couple the scintillator and fibers. Sect.~\ref{AD} describes the tests carried out at the  AD with a 70-keV antiproton beam.


\section{Detector construction}
\label{construction}

\subsection{Overall design}

\begin{figure}[ht]
\includegraphics[width=\columnwidth]{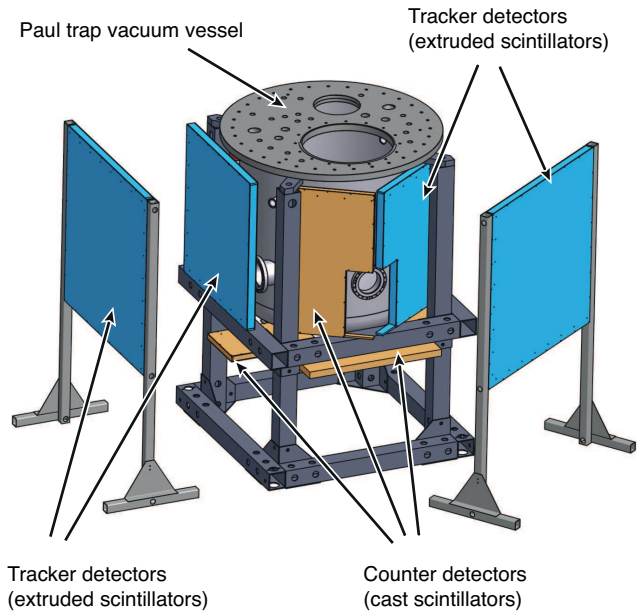}
\caption{Schematic diagram of the scintillation detector modules surrounding the Paul trap for antiprotons. Two pairs of modules (blue) consisting of 17 mm -wide extruded scintillator bars reconstruct an estimated $>15\%$ of the annihilations with 20--30 mm spatial resolution. Seven additional modules (orange) containing cast scintillators are optimized to detect the annihilations with a $>$90 $\%$ efficiency.}
\label{3Dsetup}
\end{figure}

The Paul trap is housed in a 1-m-diameter cryostat (Fig.~\ref{3Dsetup}), which has vacuum ports on its three sides comprising injection and ejection beamlines for antiprotons. The cryogenic equipment was located above the vacuum vessel. We arranged 541 scintillation bars in 11 detector modules, which together covered a solid angle of $\sim 2$$\pi$ steradians seen from the center of the trap.

\begin{figure}[ht]
\includegraphics[width=\columnwidth]{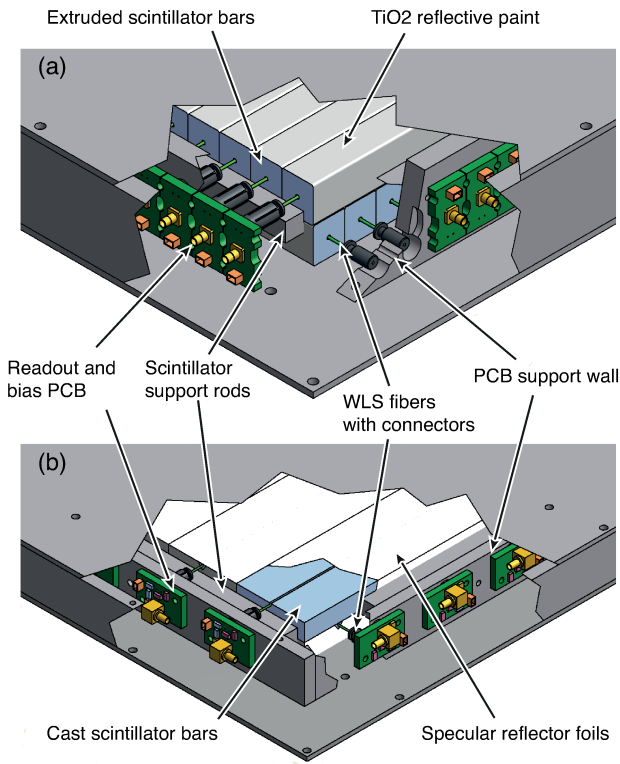}
\caption{ (a) Layout of the extruded scintillator module. A wavelength-shifting (WLS) fiber
was placed into a hole fabricated in the center of each scintillator bar. (b) Layout of the cast scintillator
module, with fibers embedded into U-shaped grooves that were machined on the surface of each scintillator bar. 
The fibers were coupled to SiPMs. Coaxial signal cables were guided from the PCB's along 
a narrow channel between the outer and inner walls of the frame,  and exited at the corner of the box.}
\label{structure}
\end{figure}

\begin{figure}[ht]
\includegraphics[width=\columnwidth]{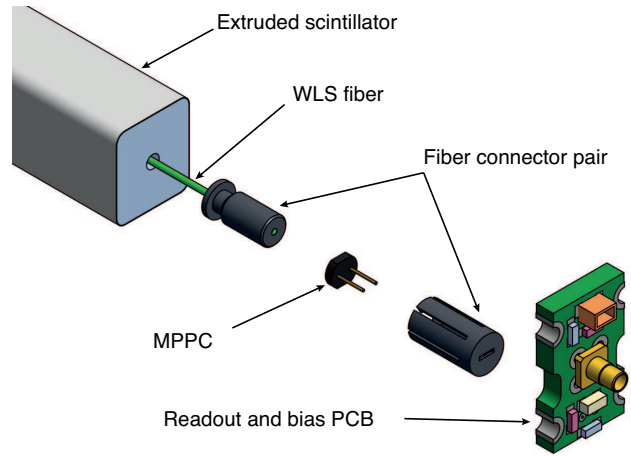}
\caption{Detail of the connection between a SiPM and WLS fiber. The fiber was glued into the male part of a connector and polished (see text). The SiPM was placed into the female counterpart, and soldered to a printed circuit board. The assembly was then bolted to the aluminium box of the module, which constituted the signal ground.}
\label{connection}
\end{figure}

Seven of the modules [Fig.~\ref{structure} (b)] ranged between sensitive areas of $0.25$$\times$$0.45$ m$^2$ and $0.8$$\times$$0.45$ m$^2$, and contained cast scintillator bars with a cross section of 50$\times$12.7 mm$^2$ which were arranged in a XY geometry. They were optimized to detect annihilations that occur in the trap with a high efficiency ($>90\%$). The four remaining modules [Fig.~\ref{structure} (a)] had larger active areas ($\sim1$$\times$$1$ m$^2$), and contained extruded scintillators with a smaller cross section of 17$\times$19 mm$^2$ which were designed for higher spatial resolution and tracking efficiency of the pions ($>15\%$). 

Each module was housed in an aluminum box with 3-mm-thick walls, with support beams which held the scintillator bars in place. Black adhesive sheets were lined along the inside the boxes, and provided the light tightness. WLS fibers embedded in each scintillator bar collected the scintillation light. The wavelength-shifted light was subsequently guided along the fibers, and coupled to SiPMs which were connected to one end of the fibers.

The detection and tracking efficiencies described above are optimized values, which were estimated using a Monte Carlo simulation based on the GEANT4 package \cite{geant4}. For this, a computer model of the trap was created, including its electrodes and vacuum chamber. We allowed antiprotons to annihilate on the electrodes according to the experimental branching ratios \cite{horicherenkov}, and tracked the secondary particles for $\sim 200$ ns, or until all the particles left the 4$\times$4$\times$4 m$^3$ area around the trap. The detection of charged pions was defined as events which deposited more than $\sim 0.4$ MeV of energy into the scintillators. 

\subsection{Cast scintillators}

The cast scintillators (Saint-Gobain Crystals BC-408) with a polyvinyltoluene (PVT) base were first milled into the desired shapes. The surfaces were then polished using a two-blade fly-cutter. This consisted of an angled forecutter with a coating of polycrystalline diamond (PCD) deposited on its edge, and a flat finishing cutter with a monocrystalline diamond (DIXI Polytools type 20370 version C) edge. The two blades were positioned on a 85-mm diameter mount which was rotated at 2000 revolutions per minute (rpm) and fed at a speed of 250 mm/min. The depth of the cut during the final pass of the tool over the scintillator was adjusted to $\sim 10$ $\mu{\rm m}$, to attain a mirror-like surface. No additional polishing of the surfaces was carried out.  A 1.2-mm-wide, 1.5-mm-deep U-shaped groove [Fig.~\ref{structure} (b)] was next cut along the surface of each scintillator \cite{groove1,bolen,uozumi,aota,kawagoe,nagano}. For this a two-flute, ball-nose slot drill made of carbide with a diamond coating applied by chemical vapor deposition (CVD) was used. The mill was rotated at 2000 rpm and fed at a speed of 160 mm/min. Each pass cut the groove in increments of 200 $\mu$m. During machining, it was essential to constantly blow dry air on the groove to avoid immediate melting of the scintillator. No other lubricants were used. After this machining, optical grease  (Eljen Technologies EJ-550) was applied to the groove, and the WLS fiber tightly pressed into it, to ensure that no bubbles formed in the contact between the fiber and scintillator. The bars were then wrapped in specular reflector foils (3M Vikuti), which have a reflectivity of $>98\%$ at visible wavelengths.

\subsection{Extruded scintillators}

A detector module containing extruded scintillators bars manufactured by Fermilab is shown in Fig.~\ref{structure} (a). Each bar with a cross section of 19$\times$17 mm$^2$ contained circular, or slightly oval, holes of diameter $d=2-3.5$ mm along its center. During the extrusion of the scintillators, TiO$_2$ diffuse reflector coatings were deposited on their surfaces \cite{pla-dalmau2001}. The attenuation length of the scintillators according to Ref.~\cite{pla-dalmau2003} was around $\sim 40$--50 mm. 

We also tested extruded scintillators manufactured by CI-Kogyo K.K, with a larger cross section of 10$\times$40 mm$^2$ 
and smaller hole diameter $1.5-2$ mm. These prototype bars were fabricated in 2010, by melting polystyrene (PS) pellets (Dow Chemicals No. 679) mixed with two kinds of fluorescent dye (PPO and POPOP of respective concentrations $\sim 1\%$ and $\sim 0.03\%$), and extruding them as in Ref.~\cite{pla-dalmau2001}. Some 0.3-mm-thick layers of TiO$_2$ were deposited on the surfaces by co-extrusion. The ends of each bar were roughly milled; polishing these surfaces or depositing layers of TiO$_2$ paint on them did not significantly increase the light yield. This may indicate the relatively short attenuation length of this scintillator.

We attempted to increase the light yield by filling optical cement (Saint-Gobain Crystals BC-600)  into the scintillator holes before inserting the WLS fibers. It was difficult to obtain a homogeneous filling, especially in the case of long ($\sim 1$ m) scintilallator bars, although shaking the scintillator helped to reduce the forming of bubbles in the cement. We also attempted to coat the fibers with optical grease (Eljen Technology EJ-550) before the insertion into the hole. Due to the high viscosity of the grease, it was similarly difficult to achieve a homogenous filling, although the viscosity could be reduced by heating the grease to a temperature $T>35$$^\circ$C.

\subsection{Fibers and silicon photomulipliers}

The WLS fibers (Kuraray Co., Ltd. Y-11(200)M) contained round polystyrene cores of refractive index $n=1.59$, which were doped with 200 parts per million (ppm) of K27 dye \cite{dyshkant}. The peak absorption and emission wavelengths were $\sim 420$ and $\sim 476$ nm. Each core was surrounded by an inner cladding made of polymethylmethacrylate ($n=1.49$), and an outer core made of fluorinated polymer ($n=1.42$). These fibers of diameters $d=1$ or $1.2$ mm had numerical apertures of $\sim 0.72$ and light attenuation lengths of $\sim 3.5$ m.

 It was essential to position the SiPMs as close as possible to the cleaved ends of the fibers and center them on the fiber axis, to attain high coupling efficiencies. For this we used plastic connectors developed by the T2K experiment \cite{kawamuko}. One of the fiber ends was glued to the male part of the connector using optical cement (Fig.~\ref{connection}). The fiber end was then polished, using polishing paper with diamond grain sizes that were progressively reduced from $10$ $\mu{\rm m}$ to $1$ $\mu{\rm m}$. The other end of the fiber was left unpolished.

To further increase the light yield, we attempted to polish the ends of the fibers opposite the SiPMs, and apply aluminum reflector layers of a few micron thickness by vacuum evaporation. We could not, however, attain a high-quality reflector surface that was precisely perpendicular to the fiber axis by hand-polishing. For 15--20$\%$ of the fibers treated in this way, a substantial amount of light leaked out of the aluminized ends when the middle sections of the fibers were illuminated.

We tested two types \cite{MPPCdevelopment} of SiPMs of active area 1$\times$1 mm$^2$, i): Hamamatsu Photonics K.K. multi-pixel photon counters (MPPC) type S10362-11-050C, which contained 
400 pixels of size 50$\times$50 $\mu$m$^2$, ii): S10362-11-025C containing 1600 pixels of size 25$\times$25 $\mu$m$^2$. The 400-pixel type had higher photon detection efficiency ($\sim 50\%$ versus $\sim 25\%$) and gain ($\sim 8\times 10^5$ versus $\sim 3\times 10^5$), due to the smaller dead area and higher capacitance of its larger pixels. These SiPMs are often used to read out, e.g., tracker scintillators which detect individual hits of minimum-ionizing particles with a high sensitivity.  The 1600-pixel SiPMs, on the other hand, had lower dark current and higher dynamic range arising from the larger number of pixels which can sustain more simultaneous photon hits. These detectors are normally used to read out scintillators in calorimetric applications. 

We employed the 1600-pixel SiPMs, since many AD experiments involve an abrupt burst of pion hits which are superimposed on a less intense background of low-rate hits. Such a distribution may be caused by a pulsed beam of antiprotons arriving in the trap, followed by the slow annihilation of antiprotons spilling out of the trap.
The recovery time needed for a fired pixel to recharge was around $\sim$20 ns \cite{renker2006}.
The typical gain, dark current rate, and Geiger breakdown voltage were respectively $\sim 3\times10^6$, $\sim$ $5\times 10^5$ Hz, and $V_0\sim70\pm1$ V, according to the specifications of the company and Ref.~\cite{MPPCperformance}. The $V_0$-value had a temperature dependence of $dV_0/dT=60$mV$/^\circ $C, and a variation of $10\%$ \cite{sakumaMPPC}. By adjusting the bias $\sim 1$ V above $V_0$, the optimal tradeoff between gain and noise characteristics was attained.

The SiPMs were placed inside the 4 mm--diameter female counterparts of the plastic connector (Fig.\ref{connection}), and soldered onto double-layered printed circuit boards made of glass epoxy (Panasonic R-1705). Two circuit boards 16.6$\times$28.5 mm$^2$ and 20$\times$33 mm$^2$ were used for the two module types shown in Fig.~\ref{structure} (a) and (b), respectively. They supplied the bias, filtering, and readout decoupling according to the circuit diagram shown in Fig.~\ref{expsetup}. The boards were firmly grounded on the aluminum box of the detector module. The output signals were transmitted using 1--1.5-m-long, SMC-type coaxial cable assemblies to the outside of the module.

\section{Pion beam measurements}
\label{T9}

\begin{figure}[ht]
\includegraphics[width=\columnwidth]{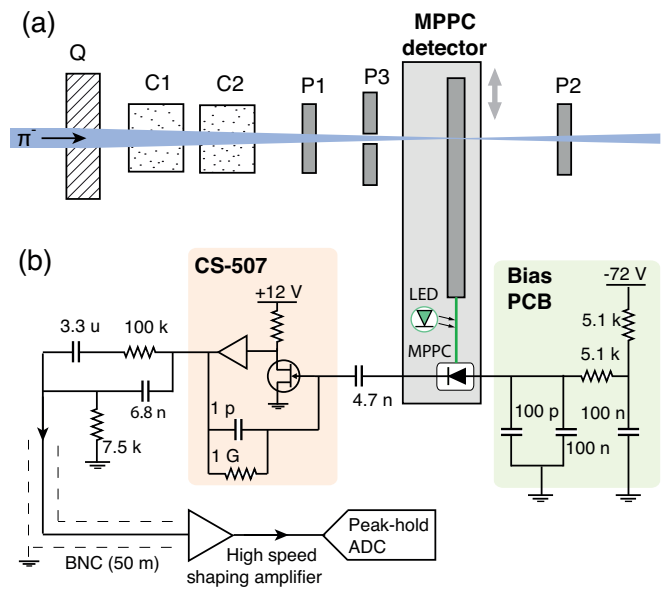}
\caption{Setup of the photoelectron yield measurement at the T9 beamline of CERN. 
The arrival of the 1 GeV/c  pions were detected by the coincidence between two scintillation counters $P1$ and $P2$. Two gas
Cherenkov detectors ($C1$ and $C2$) were used to reject the electron background. The SiPM was read out
by a charge-sensitive preamplifier. The signal was amplified by a shaping amplifier with a 50-ns time constant.
A peak-sensing ADC measured the signal height.}
\label{expsetup}
\end{figure}

The T9 beamline of CERN provided a secondary beam of pions, muons, and electrons \cite{t9zone}. It was produced by extracting protons of momentum $p=24$ GeV/c from the PS, and directing them onto a metallic target. The charge and momentum of the secondary particles were selected by pairs of dipole magnets and slits located downstream of the production target. For this experiment (Fig.~\ref{expsetup}), we used negative pions and muons with $p=0.9$--1.1 GeV/c, which is equivalent to the highest momenta of the pions emerging from antiproton annihilations \cite{horicherenkov}.

\begin{figure}[ht]
\includegraphics[width=\columnwidth]{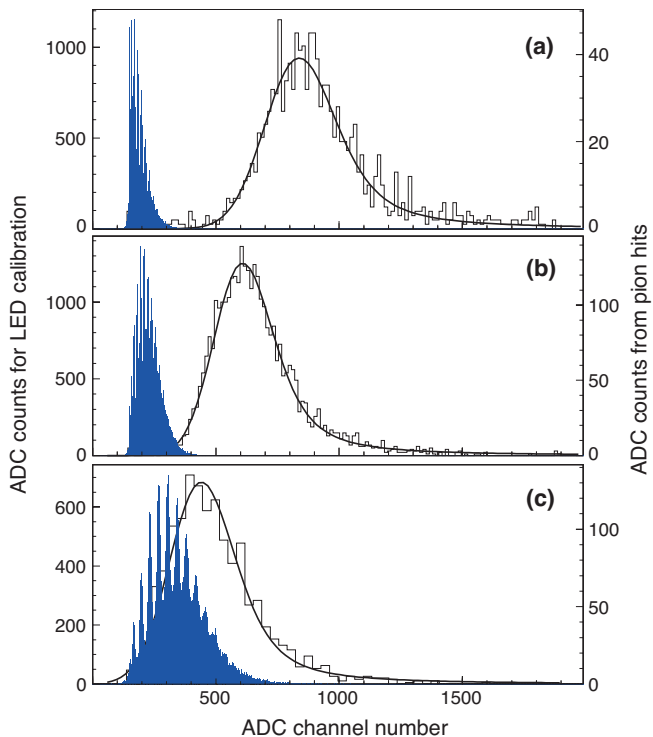} 
\caption{ (a): Pulse height distribution of BC-408 cast scintillator of 12.7 mm thickness which is struck with 1 GeV/c pions (open histogram). The response against a pulsed LED light source (blue filled histogram) is shown superimposed. Peaks corresponding to 
multiple photoelectron numbers were resolved and used to convert the ADC channels to photoelectron numbers
(see text). The solid line indicates the best fit of a convoluted Landau-Gaussian function. Similar histograms for (b) 19-mm-thick 
extruded scintillators  constructed by Fermilab and (c) 10-mm-thick ones by C.I. Kogyo K.K. and embedded with 1.2-mm-diameter WLS fibers are also shown.}
\label{result}
\end{figure}

 The secondary beam was allowed to pass through two Cherenkov counters $C_1$ and $C_2$ of length 2.5 and 5 m, which were filled with nitrogen gas at a pressure of $P\sim 1.2$ bar. The Cherenkov signals were used to reject the electron contamination in the beam. The beam then traversed two plastic scintillators $P_1$ and $P_2$, and a third scintillator $P_3$ with a 10--mm-diameter hole which defined the size of the beam. The data acquisition was triggered by events for which $\overline{C_1}\cdot \overline{C_2}\cdot P_1\cdot P_2\cdot\overline{P_3}$.

\begin{figure}[ht]
\includegraphics[width=0.85\columnwidth]{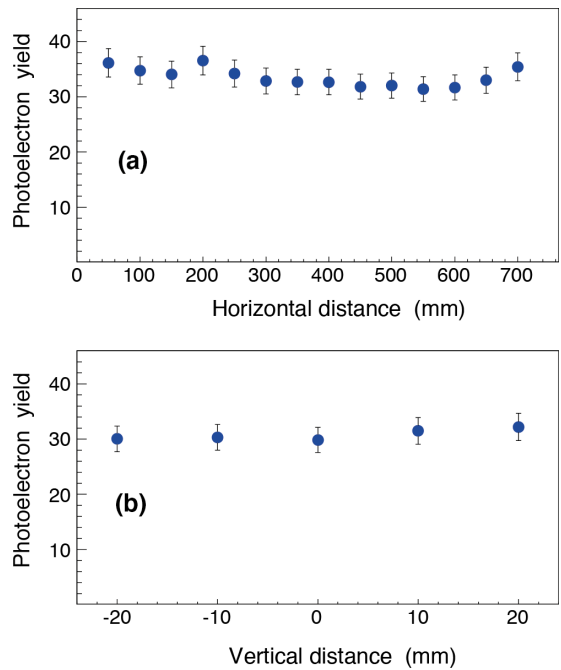}
\caption{(a) Spatial distribution of photoelectron yield $\Gamma$ of BC-408 scintillator of size 750$\times$50$\times$12.7 mm$^3$
along the axis parallel to the WLS fiber. (b) Distribution along the axis perpendicular to the fiber.
}
\label{uniformBC}
\end{figure}

\begin{figure*}[ht]
\includegraphics[width=\textwidth]{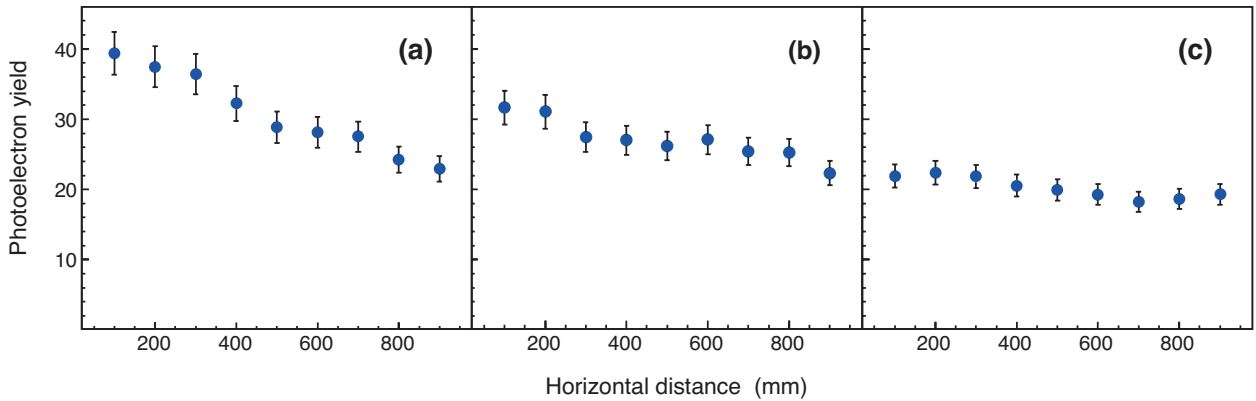}
\caption[width=\textwidth]{Spatial distributions of photoelectron yield of extruded scintillators constructed by Fermilab of size 1000$\times$17$\times$19 mm$^3$ along the axis parallel to the WLS fiber. (a) Optical grease, (b) cement, and (c) air gap were used as optical contacts between the fiber and the scintillator.}
\label{uniformFermi}
\end{figure*}

The anodes of the SiPMs were biased between $-70$  and $-73$ V, corresponding to $\sim$1 V above the Geiger breakdown voltage. The cathode signal was decoupled by a 4.7-nF capacitor, before being read out by a hybrid charge-sensitive preamplifier (Clear Pulse Co., Ltd. CS-507) with a charge-to-voltage conversion coefficient of 0.1 V/pC and integration time constant of $\sim 1$ ms. A high-pass filter then differentiated the voltage signal with a time constant of $\sim 50$ $\mu{\rm s}$ as in the design of Ref.~\cite{photocathode2005}. The preamplifier and high-pass filter were implemented on a single four-layer printed circuit board. The signal was transmitted by coaxial cables over a distance of $\sim 50$ m. It then entered a shaping amplifier (Clear Pulse 4076 custom) consisting of a passive first-order differentiator and active second-order integrator with a time constant of $\tau=50$ ns, which produced an unipolar, semi-Gaussian output pulse. The small shaping time in this final stage was needed to reduce the effects of integrating over the dark current of the SiPM, which had a typical rate of 0.1--1 MHz. The pulse amplitudes were measured by a 32-channel peak-sensing analog-to-digital converter (ADC, CAEN S.p.A. V785) with a vertical resolution of 12 bits. For data acquisition and analysis, the MIDAS \cite{midas} and ROOT \cite{root} software packages were used.

The gain of the SiPM was calibrated by measuring the light pulses produced by a light emitting diode (LED) with an emission wavelength of $\sim 420$ nm. The light intensity was adjusted so that the SiPM detected an average of 3--10 wavelength-shifted photons. In Figs.~\ref{result} (a)--(c), the distributions of signal amplitudes of LED light pulses measured in this way are indicated by the solid histograms. In SiPMs, secondary afterpulsing and optical crosstalk effects \cite{MPPCdevelopment,MPPCperformance} can cause a spurious increase in the number of discharging pixels $\lambda_{\rm meas}$, relative to the number of initial photoelectrons. These effects are strongly dependent on gain and temperature. To estimate the true number of initial photoelectrons $\lambda_{\rm real}$, we carried out a statistical analysis \cite{balagura2006} of the pulse height distributions of Figs.~\ref{result} (a)--(c). Assuming that the number of photons arriving from the LED follows a Poisson distribution, the photoelectron number can be estimated using the equation, $\lambda_{\rm real}= -\log( N_{\rm zero}/N_{\rm total})$. Here $N_{\rm total}$ denotes the total number of LED flashes, and $N_{\rm zero}$ the number of flashes in which no photon was detected. In this way, the combined afterpulsing and crosstalk probability $\varepsilon=1-\lambda_{\rm real}/\lambda_{\rm meas}$ was found to be 15--20$\%$ at a photomultiplier gain of $\sim 5\times10^6$, and the ADC channel numbers were converted to photoelectron numbers. This calibration was carried out every $\sim 10$ min during the experiment with the pion beam. 

We also estimated $\varepsilon$ using an alternative method of measuring the peak height distributions of the dark current. Most of the dark counts involved the discharge of single pixels; in $\sim 20\%$ of the cases, however, 1--3 additional
pixels fired within the $\sim 50$-ns shaping time of the amplifier, due to afterpulsing and crosstalk effects. The number of counts in these 2-pixel and 3-pixel events, relative to the total number of dark current  events, provide an estimation of $\varepsilon$. The values estimated by the two methods agreed within the experimental uncertainties.

The open histograms of Figs.~\ref{result} (a)--(c) show the pulse height distributions of the pion events measured using various scintillators. Each histogram contains typically 10$^5$ pion hits. From these data, we obtained the photon yield $\Gamma$ in the following way: first, we determined the most probable energy loss (i.e., the maximum of the distribution) by fitting a convoluted Landau--Gaussian function on the spectra \cite{balagura2006,cheshkov}. We then corrected this value for crosstalk and afterpulsing effects, using the probabilities determined using the LED calibration measurements. In the examples of Fig.~\ref{result} (a)--(c), the obtained $\Gamma$-values were respectively $36\pm 3$, $30\pm 2$, and $6\pm1$ photoelectrons. We repeated the measurements on at least 3 specimens of the same detector configurations. The experimental uncertainty on $\Gamma$ was taken as the square root of the quadratic sum of the statistical uncertainties of the three measurements, and the systematic ones of the calibration.

\begin{table*}
\caption{\label{tab:yield} Photoelectron yields $\Gamma$ of various scintillator, WLS fiber, and silicon photomultiplier 
configurations measured using the $\sim$1 GeV/c pions. Cast (type BC-408 manufactured by Saint Gobain Crystals) 
and extruded A, B (manufactured by Fermilab), and C (CI Kogyo K.K.) scintillators are compared.}
\begin{ruledtabular}
\begin{tabular}{cccccccc}
\multicolumn{3}{c}{Scintillator}    	          		                             & \multicolumn{3}{c}{WLS Fiber}          & MPPC & $\Gamma$\\
\cline{1-3}\cline{4-6}
Type  & Wid. $\times$ Thk. (mm$^2$)		&  Reflector		& Dia. (mm)	 	& Contact  & Reflector	& Pixel	&  Photoelectrons \\  
 \hline \\[1mm]
  
 Cast  &  50 $\times$ 12.7	&3M ESR	& 1.2	 	& grease	& Al		& 1600 & 37 $\pm$ 4		\\[1mm]
  			         			 &  50 $\times$ 12.7	&3M ESR	& 1.2		& grease	& --	& 1600 &37 $\pm$ 3	\\[2mm]  \\
                           
Extruded A     &  17 $\times$ 19 & TiO$_2$ 	&1.0		&grease 		& Al		& 1600 & 34 $\pm$ 3	\\[1mm]
			   &  17 $\times$ 19 & TiO$_2$ 			&1.2		&grease		& --		& 1600 & 31 $\pm$ 2	\\[1mm]
		            &  17 $\times$ 19 & TiO$_2$ 			&1.2		&grease		& Al 	& 1600 & 36 $\pm$  3	\\[1mm]
			     &  17 $\times$ 19 & TiO$_2$ 			&1.2		&glue		& Al 	& 1600 & 32 $\pm$ 3	\\[1mm]
		            &  17 $\times$ 19 & TiO$_2$ 			&1.2		& --		& Al		& 1600 & 26 $\pm$ 3	\\[2mm]\\

Extruded  B       &  40 $\times$ 10 & TiO$_2$   			&1.2		&grease		& Al		& 1600 & 14	$\pm$1 \\[1mm]
                           &  40 $\times$ 10 & TiO$_2$ 	&1.2		&grease	& Al & 400 & 19	$\pm$3	\\[2mm]  \\

Extruded  C      &  40 $\times$ 10 & TiO$_2$     &1.2   	&grease   & Al      & 1600 &   8 $\pm$ 1 \\                               
\end{tabular}
\end{ruledtabular}
\end{table*}

The highest photoelectron yields of $\Gamma=35-39$ were obtained for the cast scintillators (Table~\ref{tab:yield}). This is presumably due to the fact that scintillators with a polyvinyltoluene base generally have a technical light yield \cite{birks,moser} which is 15--25\% higher than those with a polystyrene base. Moreover the long ($\sim$2 m) light attenuation length in the BC-408 material and the high reflectivity of the specular reflector foils should in principle allow the scintillation photons to make on average 20--30 reflections inside the bar and illuminate a larger section of the WLS fiber. As shown in Fig.~\ref{result} (a), the signal is well-separated from the background of dark current counts, with a signal-to-noise ratio $>7$.

We next measured the spatial uniformity of $\Gamma$ for the cast scintillator, by varying the position of the pion beam along on its surface. In Fig. \ref{uniformBC} (a), the spatial distribution of $\Gamma$ along the horizontal (i.e, parallel to the fiber) axis of the scintillator are shown; here $x=0$ corresponds to the position of the SiPM. Fig.~\ref{uniformBC} (b) shows the distribution obtained by scanning the beam along the 50-mm-width of the bar perpendicular to the fiber axis, where $y$ denotes the distance of the beam center from the fiber. The variations in the spatial uniformity was found to be within the experimental uncertainty on $\Gamma$ of $\sim$10\%.

The $\Gamma$-values for the Fermilab extruded scintillators of 19$\times$17 mm$^2$ cross section (denoted as Extruded A in Table~\ref{tab:yield}) measured at a distance of $x=100$ mm from the SiPM was 34--38 photoelectrons. These yields were $\sim$30$\%$ less than those of cast scintillators of the same thickness, and roughly agree with the results reported by other groups \cite{pla-dalmau3,izmaylov,balagura2006}. This provides sufficient single-to-noise ratio ($>6$) above the dark current for detecting minimum-ionizing particles [Fig.~\ref{result} (b)]. Extruded scintillators with no cement or grease in the holes retained a spatial uniformity over the 1-m length of the bar which was better than $\sim 15\%$ [Fig.~\ref{uniformFermi} (c)]. Placing grease [Fig.~\ref{uniformFermi} (a)] or optical cement [Fig.~\ref{uniformFermi} (b)] in the holes increased the $\Gamma$-values near the SiPMs to $\sim$39--41 photoelectrons. At the opposite ends of the bar ($x=900$ mm), however, the light yield dropped to $\Gamma\sim$25, which is similar to values obtained without any filling. This poor uniformity appears to indicate that it is difficult to properly distribute the cement or grease along the entire length of the bar. We also measured the yield of an extruded scintillator with a cross section of $40\times10$ mm$^2$ (denoted as Extruded B in Table~\ref{tab:yield}) manufactured by Fermilab, which resulted in a value $\Gamma\sim$14--19 depending on the SiPM.

No obvious difference in $\Gamma$ was observed between WLS fibers of diameters $d=1$ mm and $1.2$ mm. This may be due to the fact that the active area of the SiPM was smaller than the fiber diameter. We used $d=1.2$ mm fibers in the final detector, according to the observation in Ref.~\cite{beznosko} that the alignment between the SiPM and fiber may become less critical for larger-diameter fibers. No significant increase in $\Gamma$ was found for the fibers with aluminum reflectors deposited on the rear ends. The reason for this is not understood, but it may be due to the insufficient quality of the reflecting surfaces.

The extruded scintillators which were manufactured by CI-Kogyo K. K. as prototypes in 2010, yielded values of $\Gamma\sim 7$--9. This was insufficient to properly separate the signal from the dark current [Fig.~\ref{result} (c)]. The reason for this is not understood, but extensive efforts to reduce the impurities and moisture, and control the temperature during the extrusion process have been made by the manufacturer since 2010. These measurements do not represent the latest such efforts.

As expected, the SiPMs containing 400 pixels of size 50$\times$50 $\mu$m$^2$ provided a $\sim$35\% increase in the $\Gamma$-values, compared to the 1600-pixel ones. We also tested detectors (Hamamatsu Photonics K. K. S10362-33-050C) with a larger active area of 3$\times$3 mm$^2$, consisting of 3600 pixels of size $50$$\times$$50$ $\mu$m$^2$, but the high dark current ($\sim 6\times 10^6$ Hz) was found to be a significant drawback in our application.

\section{Antiproton beam measurements}
\label{AD}

\begin{figure}[ht]
\includegraphics[width=\columnwidth]{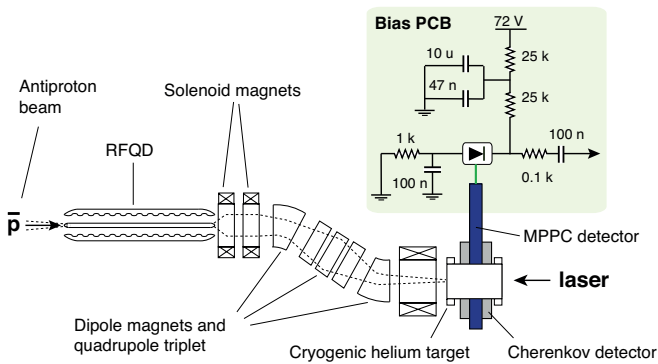}
\caption{Experimental setup to measure the response of the scintillator at the Antiproton Decelerator (not to scale). Dashed lines represent the beam envelope. A radiofrequency quadrupole decelerated the antiproton beam from a kinetic energy of $E=$5.3 MeV to 70 keV. 
A momentum analyzer transported the 70-keV antiprotons to the position of a helium gas target. Charged pions emerging from antiproton annihilations in the target and beamline walls were detected by an acrylic Cherenkov counter and scintillator.}
\label{adsetup}
\end{figure}

Fig.~\ref{adsetup} shows the experimental setup which was used to measure the response of the scintillators against antiproton annihilations. The AD provided a 300-ns long pulsed beam containing $(2-3)$$\times$10$^7$ antiprotons with a kinetic energy $E=5.3 $MeV at a repetition rate of $0.01$Hz \cite{mhori2003,mhori2006}. This beam was allowed to pass through a radiofrequency quadrupole decelerator (RFQD), which reduced the energy of some $\sim 30\%$ of the antiprotons to $E=70$ keV. The decelerated antiprotons were diverted by an achromatic momentum analyzer which was connected to the exit of the RFQD, and focused into an experimental helium target. The analyzer consisted of two dipole magnets which deflected the beam at an angle $\Theta = 20^{\circ}$, and three 1-T solenoid magnets. The beam profile was measured using microwire secondary electron emission detectors \cite{photocathode2005,ppac2004,linac4}. Some $\sim$70\% of the antiprotons emerged from the RFQD without being decelerated, and annihilated on the walls of the analyzer.

\begin{figure}[ht]
\includegraphics[width=\columnwidth]{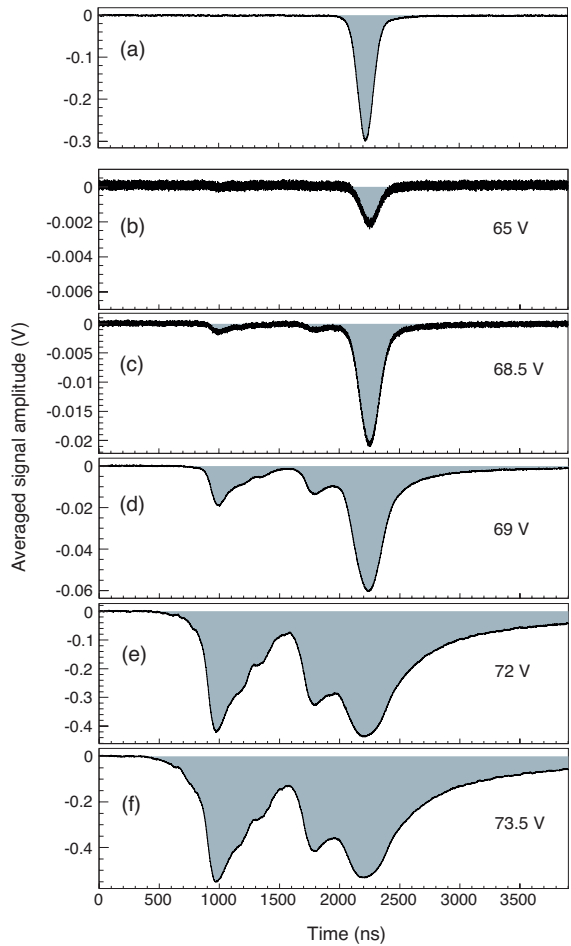}
\caption{Output signals of (a) acrylic Cherenkov detector and (b)--(f) scintillators read out by WLS fibers and SiPMs
biased at various voltages $V_{\rm bias}$. Each waveform is an average of 14--15 antiproton pulses arriving at the experimental target.}
\label{chercompare}
\end{figure}

\begin{figure}[ht]
\includegraphics[width=\columnwidth]{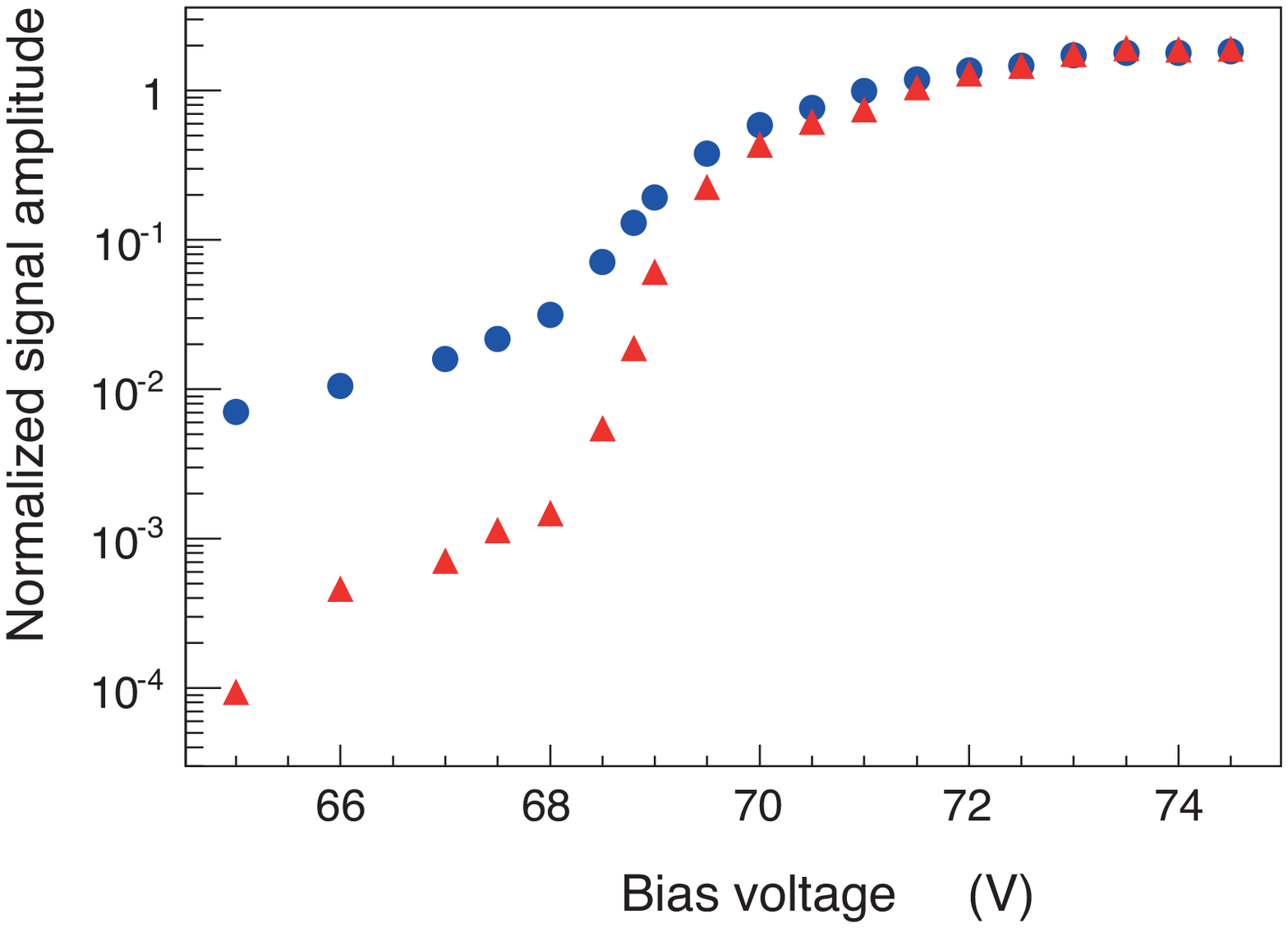}
\caption{Normalized amplitudes of the output signals of the scintillator against the pulsed AD beam,
plotted against the bias voltage applied to the SiPM. The intensity of the peaks at $t=950$ ns (filled triangles) and $2250$ ns (circles) in Fig.~\ref{chercompare} are shown superimposed. The signal amplitude increases rapidly around the Geiger breakdown voltage $\sim 69$ V, and then saturates at higher bias voltages, presumably because the charge in the SiPM is depleted.}
\label{saturation}
\end{figure}

A Cherenkov detector \cite{horicherenkov} made of UV-transparent acrylic (Mitsubishi Rayon, Acrylite000) of size $300$$\times$100$\times$20 mm$^3$ was placed at a distance of $\sim 1$ m from the experimental target. Some of the pions that emerged from the antiproton annihilations traversed the detector, and the resulting flash of Cherenkov light was detected by a fine-mesh photomultiplier (Hamamatsu R5505GX-ASSYII). This photomultiplier had a photocathode of diameter $17.5$ mm, and operated at a gain of $5\times 10^4$. This detector has been used for many years in precision laser spectroscopy of antiprotonic helium atoms \cite{nature_2011,mhori2006}, and its linear behavior against high fluxes of pions have been extensively characterized \cite{horicherenkov}. The analog waveform of the photomultiplier was recorded using a digital oscilloscope.

In Fig.~\ref{chercompare} (a), the waveform of the Cherenkov signal taken as an average of 14 pulses, each containing $\sim 6\times 10^6$ antiprotons arriving in the target, is shown. The instantaneous flux of pions was so high that individual annihilations could not be resolved. The 300-ns-long Gaussian-shaped signal at $t\sim 2250$ ns corresponds to the envelope of the pulsed beam of antiprotons 
which annihilated at the target position. The annihilations that occurred at other positions along the RFQD and analyzer cannot be detected, because the resulting pion flux is too low compared to the sensitivity of the detector.

We also placed a cast scintillator of size $750$$\times$50$\times$10 mm$^3$, which was read out by a WLS fiber and a 1600-pixel SiPM, at the same position. The cathode of the SiPM was biased at $V_{\rm bias}=65.0-73.5$ V which corresponded to a voltage between $-4$ and $+4.5$ V around the Geiger breakdown value $V_0$. The anode was grounded using a 1-k$\Omega$ resistor and 100-nF capacitor, similar to the configuration in Ref.~\cite{tript}. The anode signal was decoupled by a 100-nF capacitor, and the waveform recorded by a digital oscilloscope of input impedance $50$ $\Omega$. In Fig.~\ref{chercompare} (b), the signal measured at a low bias of $V_{\rm bias}\sim 65$ V corresponding to $V_0-4$ V is shown. The detector operated as a conventional avalanche photodiode of low gain. As expected, a single peak of amplitude $\sim 2$ mV appeared at $t\sim 2250$ ns, as in the Cherenkov counter case [Fig.~\ref{chercompare} (a)]. 

As $V_{\rm bias}$ neared the breakdown voltage [Fig.~\ref{chercompare} (c), $V_{\rm bias}=68.5$ V], however, two secondary peaks at $t\sim 950$ and 1800 ns appeared, while the signal amplitude increased by an order of magnitude, to $\sim 20$ mV. 
This was caused by the higher sensitivity and gain of the SiPM, which now detected the weak scintillation light that arose from the annihilations in 
the two bending magnets located upstream of the target (Fig.~\ref{adsetup}). As we exceeded the breakdown voltage 
[Fig.~\ref{chercompare} (d), $V_{\rm bias}=69$ V], many more structures appeared which corresponded 
to the small number of antiprotons that scraped the walls of the analyzer or the secondary electron emission monitors.  
The detector response against the peak at $t=2250$ ns of amplitude $\sim 60$ mV was clearly saturated. Between $V_{\rm bias}=72$ [Fig.~\ref{chercompare} (e)] and 73.5 V
[Fig.~\ref{chercompare} (f)], the signal amplitude ($\sim 400$ mV) no longer increased, presumably because the charge stored in the SiPM was depleted. 
The waveform was also deformed by the recharging time of the bias circuit and SiPM.
When compared with the measurements in Sect.~\ref{T9}, we estimated that for a setting of $V_{\rm bias}\sim V_0+1.5\sim 70.5$ V, which was an optimal compromise between gain and dynamic range, around $\sim 10$ simultaneous pion hits would saturate the SiPMs. 

In Fig.\ref{saturation}, the amplitudes of the two intense peaks at $t=950$ ns (filled triangles) and $2250$ ns (circles) normalized to the antiproton beam intensity measured by the Cherenkov counter are shown for various values of $V_{\rm bias}$. The rapid increase at $\sim$68.8 V is due to the SiPM reaching Geiger breakdown. From these results, we decided to operate the scintillator in two modes, i): at a bias of $V_{\rm bias}\sim V_0-4\sim 65$ V to measure the timing profile of the antiproton beam in a relatively linear way, ii): at $V_{\rm bias}\sim V_0+1.5\sim 70.5$ V to detect single pion hits.

\section{Discussions and conclusions}
\label{conclusion}

In conclusion, we constructed a 541-channel, segmented scintillator for detecting and tracking charged pions emerging from antiproton annihilations. Both cast and extruded scintillators of thicknesses $t_d=12.7$--19 mm were used, which were read out by wavelength-shifting (WLS) fibers and silicon photomultipliers (SiPMs). The design was optimized to attain high photoelectron yields of around $\Gamma=35-38$ for minimum-ionizing particles. This was sufficient to detect antiproton annihilations in a future Paul trap with an estimated efficiency of $>90\%$. An important source of background was the dark current of the SiPMs, which reached a maximum amplitude equivalent of 3--4 simultaneous pixel discharges within the $\sim 50$-ns shaping time of the amplifier. This implied that a $\Gamma$-value of $>20$ was needed to achieve a sufficient signal-to-noise ratio. In the current detector design, it may therefore be difficult to significantly reduce the thicknesses of the scintillators (e.g., $t_d\ll10$ mm) and increase the spatial resolution, without sacrificing the detection efficiency. The response of the scintillator against the high-intensity flux of pions emerging from the annihilation of a pulsed beam of antiprotons was studied. At low bias voltages of the SiPM below the Geiger breakdown threshold, the detector accurately measured the envelope of the pulsed antiproton beam, but
the sensitivity was too low to detect individual annihilations. At higher bias voltages, the detector became sensitive to individual pion hits, but a nonlinear behavior was seen.

The scintillators were recently used to detect the small number of annihilations that occurred when an antiproton beam of kinetic energy $\sim 130$ keV was allowed to traverse thin target foils \cite{pbaranni}. This technique may be used in the future to determine the total annihilation cross-sections of antiprotons on various metal targets
in the low-energy region. 

\section{Acknowledgments}

 We are deeply indebted to A.~Pla-Dalmau for supplying the extruded scintillators used in this work and for fruitful discussions. We thank the ASACUSA collaboration, R.~Dumps, L.~Gatignon, A.~L\'aszl\'o, A.~Minamino, T.~Nakaya, K.~Niita, T.~Schneider,  and P.~Zal\'an for their help. This work was supported by the European Research Council (ERC-Stg), European Science Foundation (EURYI), Monbukagakusho (grant no. 20002003), and the Hungarian Research Foundation (K103917).



 \end{document}